\documentclass[
reprint,
amsmath,amssymb,
aps,
superscriptaddress
]{revtex4-2}
\usepackage{graphicx}
\usepackage{dcolumn}
\usepackage[dvipsnames, usenames]{xcolor}
\usepackage{bm}
\usepackage[colorlinks=true,
            linkcolor=blue,
            citecolor=blue,
            urlcolor=blue]{hyperref}
\usepackage[mathlines]{lineno}
\usepackage{braket}
\usepackage{comment}

\begin{document}

\preprint{APS/123-QED}

\title{Emergent scalar field dynamics in a cosmological spacetime from GFT quantum gravity}

\author{Roukaya Dekhil}
\email{roukaya.dekhil@philosophy.su.se}
\affiliation{Nordita, KTH Royal Institute of Technology and Stockholm University,
Hannes Alfvéns väg 12, SE-106 91 Stockholm, Sweden}
\author{Federico Greco}
\email{federico.greco@pd.infn.it}
\affiliation{
Dipartimento di Fisica e Astronomia “G. Galilei”, Universit\`a degli Studi di Padova, via Marzolo 8, I-35131 Padova, Italy}%
\affiliation{
INFN, Sezione di Padova, via Marzolo 8, I-35131 Padova, Italy
}%
\author{Stefano Liberati}
\email{liberati@sissa.it}
\affiliation{SISSA, Via Bonomea 265, 34136 Trieste, Italy}
\affiliation{
INFN sezione di Trieste, Via Valerio 2, 34127 Trieste, Italy}
\affiliation{IFPU, Institute for Fundamental Physics of the Universe, \\
via Beirut 2, 34014 Trieste, Italy}%
\author{Daniele Oriti}
\email{doriti@ucm.es}
\affiliation{Departamento de F\'{i}sica Te\'orica, Facultad de Ciencias F\'isicas,\\ Universidad Complutense de Madrid, \\ Plaza de las Ciencias 1, 28040 Madrid, Spain, EU
}


\begin{abstract}
%
We derive an effective scalar field theory for matter in group field theory condensate cosmology, starting from the fundamental quantum-gravity dynamics in a fully relational framework and encompassing both early- and late-universe regimes. The collective hydrodynamics of the underlying quantum geometry allows us to reconstruct both the homogeneous cosmological dynamics of matter and geometry and an inhomogeneous local field-theory description. Localization in space and time is defined relationally with respect to a material reference frame. At the homogeneous level, we obtain a modified scalar field theory on the emergent FLRW spacetime selected by the condensate. It recovers the standard dynamics of a massless scalar field in the late-time general-relativistic regime while retaining quantum-gravity corrections near the cosmological bounce. At the perturbative level, scalar inhomogeneities obey an effective wave equation that carries signatures of the underlying quantum-gravity microstructure. In the early-universe regime, this equation exhibits a modified dispersion relation with both dispersive and dissipative contributions. These corrections provide a concrete avenue for identifying phenomenological signatures of quantum gravity directly from a fundamental quantum-gravity framework.
\end{abstract}

\maketitle

\textbf{\textit{Introduction}} --- 
Modern cosmology rests on two pillars: General Relativity (GR) for gravity and Quantum Field Theory (QFT) for matter and other interactions \cite{maggiorebook, primordialcosmology,Wald:106274,qft2,QFT91199263, Mukhanov:991646}. Yet, despite their predictive power within their respective domains of validity, and their mathematical richness, they remain fundamentally incompatible. While GR predicts its own breakdown at cosmological singularities, QFT resists any extension to a complete description of a dynamical quantum spacetime, thus failing to incorporate the main lesson of gravitational physics. This tension motivates the search for a theory of quantum gravity (QG), explaining the emergence of spacetime, geometry, and matter from more fundamental microscopic degrees of freedom \cite{QGKiefer, Kiritsis:2019npv_string, Giesel:2012ws}.

Among the various approaches to QG, Group Field Theories (GFTs) provide a framework in which continuum spacetime can emerge collectively from the dynamics of discrete quantum-geometric entities. In particular, GFT condensate cosmology describes cosmological spacetimes as condensate states of a large number of GFT quanta~\cite{Oriti2015,Gielen_2013,Gielen2020}. In the relational formulation of the theory, massless scalar fields define physical rods and clocks, allowing one to extract the dynamics of cosmological observables without referring to a pre-existing spacetime manifold ~\cite{Marchetti_2021,Gielen2018,H_hn_2021,Li_2017, Giesel_2021_dust, marchetti2025relationalobservablesgroupfield}. The resulting effective dynamics reproduces a nonsingular bouncing cosmology at early-times and approaches, at late-times, a spatially flat FLRW universe sourced by a massless scalar field~\cite{Oriti2016,Marchetti_2022, Gielen:2013naa}.

While the emergent geometric sector of GFT condensate cosmology has been extensively investigated, the effective dynamics of matter fields on such emergent backgrounds, especially beyond the homogeneous sector, remains less developed (but see \cite{Gielen2019, Gielen:2025jcb, Marchetti:2021gcv}). This is a key step for phenomenology: cosmological observations are ultimately phrased in terms of matter degrees of freedom and their perturbations propagating on an effective spacetime. Understanding how their dynamics is reconstructed from the same microscopic condensate data that determine the geometry is therefore essential for connecting GFT cosmology with possible observational signatures of QG \cite{Addazi:2021xuf_pheno_qg}. More generally, it is crucial to extract the usual language of effective field theory from any fundamental quantum gravity formalism.

In this work, we derive the effective field theory of a scalar matter degree of freedom directly from relational GFT condensate observables, both at the homogeneous and (perturbatively) inhomogeneous levels. The main novelty is twofold. First, at the background level, we reconstruct the scalar-field equation of motion from the dynamics of the condensate phase and show how it can be interpreted as a modified scalar-field dynamics on the curved FLRW background selected by the condensate. This formulation reproduces the standard massless behavior in the late-time general-relativistic regime while retaining quantum-gravity corrections closer to the bounce. Second, at the inhomogeneous level, we derive an effective wave equation for scalar-field perturbations, whose coefficients are fixed by the underlying condensate background. In the coupled density-phase regime of the condensate, this equation contains higher-spatial-derivative and mixed derivative terms, leading to a modified dispersion relation with both dispersive and effective dissipative contributions. These corrections originate from the microscopic GFT condensate structure and provide a direct imprint of quantum-geometric effects on emergent matter propagation.

\textbf{\textit{GFT condensate cosmology}} --- 
The GFT model that we consider is defined by a field operator $\hat \varphi: {\rm{SU}}(2)^4 \times \mathbb{R}^5 \rightarrow \mathbb{C}$ which we denote $\hat \varphi(g_I,\boldsymbol{\psi})$ \cite{FreidelGFT, GFT_thomas, Oriti:2011jm}. 
The group elements $g_I,\,\,I=1,\dots4$, encode the quantum geometric degrees of freedom, while $\boldsymbol{\psi}=(\chi^\mu,\tilde \pi_\phi)$, with $\mu=0,\dots,3$, denotes four massless free scalar fields coupled to the (discrete) quantum geometry and $\tilde \pi_\phi$ is interpreted as the momentum of the fifth field $\phi$. 
The four fields $\chi^\mu$ are used as a relational reference frame with respect to which physical observables are localized \cite{Li_2017, Tambornino_2012_relational_classic}, while $\phi$ is assumed to dominate the energy-momentum budget of the classical GR system that we aim to reproduce in some regime \cite{Oriti:2016qtz, Gielen:2016dss_reviewcondensate}.



From the standpoint of the underlying quantum geometry, being a second quantized formulation of simplicial geometry \cite{Oriti:2011jm} and loop quantum gravity \cite{Oriti:2013aqa}, upon imposing diagonal invariance, namely $\hat \varphi (h\,g_I,\boldsymbol{\psi})=\hat \varphi (g_I,\boldsymbol{\psi}), \, h\ \in {\rm{SU}}(2)$, GFT quanta can be interpreted as quantum tetrahedra carrying matter degrees of freedom. Within the mean-field approximation, cosmological continuum geometries are then described by coherent condensate states $|\sigma\rangle$, which are eigenstates of the field operator with eigenfunction $\sigma (g_I, \boldsymbol{\psi})$ \cite{Oriti2016GroupGravity,Oriti:2017GFTManyBodySystem,Oriti:2021GFTCondensateEmergent}.
The latter admits a Peter-Weyl decomposition on ${\rm SU}(2)$ whose modes are labeled by irreducible spin representations $j$ \cite{gelfand_representation}. The condensate wave function is thus characterized by $\sigma_j(\boldsymbol{\psi})$.
Furthermore, we assume that  $\sigma_j(\boldsymbol{\psi})$ factorizes into a function sharply peaked around $\boldsymbol{\psi}_0=(x^\mu, \pi_\phi)$ and a \textit{reduced} wave function $\tilde \sigma_j(\boldsymbol{\psi}_0)$. 
The peaking values $x^\mu$ define relational rods and clock, while the reduced condensate wave function $\tilde\sigma_j(\boldsymbol{\psi}_0)$ captures the collective \textit{hydrodynamics} of the system \cite{Marchetti_2022, Gielen:2013naa, Marchetti_2021}. Its domain can be shown to be isomorphic to minisuperspace, or equivalently the space of continuum metrics (and matter fields) at a manifold point \cite{Gielen2014, Jercher_2022cosmo}.  The GFT dynamics of the model is encoded in the action $S[\hat\varphi, \hat\varphi^\dagger]= K[\hat\varphi, \hat\varphi^\dagger]+ V[\hat\varphi, \hat\varphi^\dagger]$.
We will work in a regime of negligible GFT interactions, thus with
\begin{equation}
S=\int{\rm{d}}\Omega{\rm{d}}\Omega^\prime\varphi^\dagger\left(\Omega\right) \mathcal{K}\left(g_I, g_I';\left(\chi-\chi^{\prime}\right)^2,\tilde \pi_\phi^2\right) 
\varphi\left(\Omega'\right)\,,
\end{equation}
where we defined $\Omega:=(g_I, \boldsymbol{\psi})$. We keep the form of the action rather generic (within the quantum geometric class of GFT models). The mean field dynamics for the condensate implies that the equations of motion are satisfied on average, namely $\langle \delta S[\varphi, \varphi^\dagger]/\delta \varphi^\dagger \rangle_\sigma=0$, where $\braket{}_{\sigma}$ indicates the expectation value taken with respect to the coherent state $|\sigma \rangle$.
This leads, for the peaked condensate states introduced earlier, to the following effective evolution equation for $\tilde\sigma_j(\boldsymbol{\psi}_0)$
\begin{equation}
\label{effective_GFT_EOM}
     \partial_0^2\tilde{\sigma}_j (\boldsymbol{\psi}_0)-i \gamma \partial_0\tilde{\sigma}_j(\boldsymbol{\psi}_0)-E\left(\pi_\phi\right) \tilde{\sigma}_j(\boldsymbol{\psi}_0)+\alpha^2\nabla^2 \tilde{\sigma}_j(\boldsymbol{\psi}_0)=0\,,
\end{equation}
where $\partial_0$ denotes the derivative with respect to the clock $x^0$, and $\nabla^2= \sum_{i=1}^3 \partial_i^2$ is the Laplacian with respect to rods $x^i$. The parameters $E(\pi_\phi)$ and $\gamma$ encode the form of the kinetic kernel of the action and the peak width parameters \cite{Marchetti:2021gcv}. Note that $\alpha^2$ is in general a complex variable with real and imaginary parts $\alpha_r$ and $\alpha_i$. 
We adopt a Madelung representation in terms of density and phase, writing $\tilde \sigma_j(\boldsymbol{\psi}_0)=\rho_j(\boldsymbol{\psi}_0)\, e^{i \theta_j(\tilde{\boldsymbol{\psi}}_0)}$. 

Since we are interested in cosmological configurations, we further decompose the density and phase into a dominant homogeneous background and small inhomogeneous perturbations (with localization defined relationally with respect to the material frame): 
\begin{equation}
\begin{split}
     \rho_j(\boldsymbol{\psi}_0)= \rho_{0,j} (x^0, \pi_\phi)+ \delta \rho_j(\boldsymbol{\psi}_0)\,, \\
    \theta_j(\boldsymbol{\psi}_0)= \theta_{0,j}(x^0, \pi_\phi)+ \delta \theta_j (\boldsymbol{\psi}_0)\,.
\end{split}    
\end{equation}
In the free theory we are considering, the dynamics singles out a dominant mode whose occupation number grows exponentially faster than that of all other modes \cite{Oriti2016}.
Therefore, we restrict our analysis to this mode, and we omit the label $j$ in the following. Throughout, the subscript 0 will denote homogeneous background quantities. 
Starting from the underlying quantum geometric Hilbert space, or directly in terms of (condensate) hydrodynamic averages, we can compute the relevant background observables:
\begin{align}
\label{volumeoperator}
V &:= \langle \hat V \rangle_\sigma
\\
&:= \Big\langle
\int {\rm d}g_I\,{\rm d}g'_I\,{\rm d}\boldsymbol{\psi}\,
\hat\varphi^\dagger(g_I,\boldsymbol{\psi})
V(g_I,g'_I)
\hat\varphi(g'_I,\boldsymbol{\psi})
\Big\rangle_\sigma
\simeq v\,\rho_0^2\,,
\notag
\\[1ex]
\label{eq:scalarfield}
\phi_0 &:= \langle \hat \Phi \rangle_\sigma
\\
&:= \Big\langle
\int {\rm d}\Omega\,
\hat\varphi^\dagger(\Omega)\,
\partial_{\tilde\pi_\phi}\hat\varphi(\Omega)
\Big\rangle_\sigma
\simeq \rho_0^2\,\partial_{\pi_\phi}\theta_0\,,
\notag
\end{align}
where $V(g_I,g^\prime_I)$ is the second-quantized version of the LQG volume operator~\cite{Oriti:2013aqa} and $v$ denotes the eigenvalue of the volume operator associated with the dominant spin label $j$.
In the remainder of this letter, we will also suppress the domain of the relevant fields whenever no confusion arises.


\textbf{\textit{Geometric background dynamics}} --- 
At the background level, Eq.\eqref{effective_GFT_EOM} reduces to two coupled equations for the GFT density $\rho_0$ and the phase $\theta_0$ \cite{Marchetti:2021gcv}:
\begin{align}
\ddot\rho_0-\left[\left(\dot\theta_0\right)^2+\eta-\gamma \dot\theta_0\right] \rho_0=0\,, \label{eq:density eom}\\
\ddot\theta_0+\frac{\dot\rho_0}{\rho_0}\left(2 \dot\theta_0 -\gamma \right)-\beta=0\, ,\label{Eq:phase_eom}
\end{align}
where $\beta\propto\alpha_i$~\cite{Marchetti_2022}.
The decoupled perturbative regime is obtained by setting $\alpha_i\simeq0$, which in turn implies $\beta\simeq0$. It has been shown in\cite{Marchetti_2022}, that the late-time cosmological dynamics reproduces the GR behavior only for negligible $\beta$. Nevertheless, perturbative corrections at first order in $\beta/\eta$ can be consistently incorporated without spoiling this late-time background dynamics. 

The solutions for the density and phase reduce to
\begin{equation}
\label{scalefactor_solution}
 \rho_0=\Bigg(-\frac{\mathcal{E} }{2 \mu ^2}+A e^{2 \mu   x^0} +B e^{-2 \mu   x^0}\Bigg)^{1/2},
\end{equation}
\begin{equation}
\label{solution_theta_zero_final}
    \begin{split}
        \theta_0= \frac{\gamma}{2} x^0  +  \tan^{-1} \left[ \frac {\mu }{Q } \left(2\, A \,{\rm{e}}^{2 \mu  x^0} - \frac{{\cal{E} }}{2 \mu  } \right) \right] +c 
    \end{split}
\end{equation}
where $\mu^2=\eta^2-\gamma^2/4$ and  $A$, $B$ and $c$ are integration constants, whereas $\mathcal{E}$ and $Q$ are two constants of motion \cite{Marchetti_2021}.
We restrict to the branch where $Q>0$ in Eq.\eqref{solution_theta_zero_final}, as this is also necessary to recover the GR dynamics at late-times. 

Let us now evaluate the expectation values of the volume and the scalar field operators defined in Eq.\eqref{volumeoperator} on the above defined solutions, and derive their effective dynamics. 
At late (relational) times, for the universe volume, one recovers an FLRW universe sourced by two massless scalar fields, $\phi_0$  and $x^0$ (clock), with canonical momenta satisfying $p_\phi \gg p_{x^0}$, in harmonic gauge, $N=a^3$, with $N$ being the lapse function and $a$ the scale factor. This is consistent with the fact that we use a massless free scalar field as a clock.
The corresponding physical time coordinate is $X^0=\kappa\, x^0$, with mass dimension $-1$, implying $[\kappa]=-2$.
This GR limit is obtained upon the following identifications
\begin{align}
\label{Matching_GR_row_1}
\pi_\phi &:= \sqrt{3\pi G_N}\,\bar\pi_\phi\,,
&
\mu &= \pi_\phi\,,
\\
\label{Matching_GR_row_2}
Q &= \frac{\pi_\phi^2}{\sqrt{3\pi G_N}}\,,
&
\partial_{\pi_\phi} c &= 0\,,
\end{align}
where $G_N$ is Newton's constant and $\bar \pi_\phi := p_\phi/p_{x^0}$.
On the other hand, at early-time, the singularity is resolved by a quantum bounce and the dynamics of the field deviates from the GR one, reproducing instead (from a full QG dynamics) the effective dynamics of loop quantum cosmology). For details, see \cite{Marchetti:2021gcv}.

\textbf{\textit{Effective background scalar field dynamics}} --- 
To derive the effective dynamics of the background scalar field $\phi_0$, we invert Eq.\eqref{eq:scalarfield} to express $\partial_{\pi_\phi}\theta_0$ in terms of $\phi_0$ and $\rho_0$, i.e., $\partial_{\pi_\phi}\theta_0=\phi_0/\rho_0^2$. Since $\rho_0^2\neq 0$ throughout the evolution, consistently with the nonsingular bouncing dynamics of the model \cite{Pithis:2019tvp_rev_condensate, Oriti:2021GFTCondensateEmergent}, this inversion is always well-defined. Moreover, since the solutions are sufficiently regular, the mixed derivatives commute, we can differentiate the equation of motion for $\theta_0$ in Eq.\eqref{Eq:phase_eom} with respect to $\pi_\phi$ and rewrite the resulting expression in terms of the cosmological observables, namely the volume and $\phi_0$. This leads to the effective dynamics of the background scalar field
$\phi_0$ 
\begin{equation}
\label{effective_eom_scalarfield_background}
    \ddot \phi_0- 3 {\cal{H}} \dot \phi_0- 3  {\cal{\dot H}} \phi_0 + J_0=0\,,
\end{equation}
where ${\cal{H}}:= (2 \dot \rho_0)/(3 \rho_0)$ is the relational Hubble rate. The term $J_0(x^0,\pi_\phi)$ is defined as 
\begin{eqnarray}
     J_0(x^0,\pi_\phi) := 3 \,Q\, \partial_{\pi_\phi} {\cal{H}} .
\end{eqnarray}
This source term is reminiscent of a noise-like contribution induced by the interaction between the effectively reconstructed scalar field and the underlying quantum-gravity microstructure, along the lines suggested, for instance, in~\cite{Perez:2017krv,Pellecchia:2026pwb}. Source-like terms of a similar nature have also been found in a different GFT setting in~\cite{Jercher:2023kfr,Jercher:2023nxa}, where they arise in the effective dynamics of perturbations of both a scalar field and a curvature-like variable.
In the late-time regime, this equation reduces to
\begin{equation}
    \ddot \phi_0 - 2 \mu  \dot \phi_0+ 2 Q \partial_{\pi_\phi} \mu \simeq 0\,,
\end{equation}
whose general solution is given by
\begin{equation}
\begin{split}
    \phi_0  &\simeq \frac{Q \partial_{\pi_\phi }\mu}{2 \mu} x^0+ c_1 \frac{{\rm{e}}^{2 \mu x^0}}{2 \mu}\,,
    \end{split}
\end{equation}
where $c_1$ is an integration constant. 
Imposing $c_1=0$ yields the expected late-time classical behavior, namely $\ddot \phi_0=0$ \cite{harmonic_gauge, Marchetti:2021gcv}. This condition coincides with the second relation in Eq.\eqref{Matching_GR_row_2} and is tied to the Hubble-friction-like contribution, with opposite sign, appearing in Eq.\eqref{effective_eom_scalarfield_background}. This  term is crucial for recovering the correct late-time limit. At the same time, however, it gives rise to an \textit{extensive} branch of solutions, scaling with the volume, which can be traced back to the definition of the scalar field operator in Eq.\eqref{eq:scalarfield}. 
As a consequence, the physically admissible configurations form only a subset of the full space of solutions of Eq.\eqref{effective_eom_scalarfield_background}. 
An analogous issue arises at the level of perturbations, as discussed in the following. 
An intensive definition of the GFT scalar-field operator has been proposed in \cite{Marchetti:2024nnk}; exploring whether this provides a systematic way to isolate the physical sector of scalar-field expectation values is left for future work.


We now recast the effective equation of motion as an emergent scalar-field dynamics on a curved spacetime background. There are two ways to proceed. We could start from the standard relativistic action for a homogeneous scalar field on an FLRW spacetime with arbitrary lapse, and then choose the lapse so that the variational equation reproduces Eq.\eqref{effective_eom_scalarfield_background}. The second is to fix the effective FLRW background to harmonic gauge from the outset, as suggested by the reconstructed late-time GR limit discussed in the previous section, and to encode the deviations from the standard dynamics in a modified scalar-field action.

Let us first start by assuming the standard relativistic action for a homogeneous scalar field $\phi_0$ in the cosmological symmetry-reduced sector of GR, namely
\begin{equation}
S = V_0 \int_{\mathbb{R}} dX^0 a^3 \left[ \frac{\phi_0^{\prime \,2}}{2N} - N U(\phi_0) \right],
\end{equation}
where $V_0$ is the volume of a fiducial spatial cell, $N$ is the lapse function, $X^0$ is the time parameter, a prime denotes differentiation with respect to $X^0$, and $U$ is a generic potential. Reproducing Eq.\eqref{effective_eom_scalarfield_background} from this action requires, in addition to a suitable choice of potential, fixing the lapse to $N=a^6$. This would define an effective spacetime as seen by our scalar field. However, such a spacetime is not compatible with the late-time regime reconstructed in the previous section, where a massless scalar field propagates on an FLRW background written in harmonic gauge.

For this reason, we follow the second route. We fix the lapse of the effective background to the harmonic one, and allow instead for a modification of the scalar-field kinetic structure. This realization is not unique. A simple possibility is to consider the symmetry-reduced minisuperspace action
\begin{equation}
S=V_0\int {\rm{d}} X^0 \left[ \frac{1}{2}  {\rm{e}}^{-\varsigma}  \,\phi_0^{\prime \, 2}  -U(\phi_0) \right].
\end{equation}
The equation of motion for $\phi_0$ is reproduced by choosing $\varsigma=\ln(V/v)$.
The corresponding potential is
\begin{equation}
U(\phi_0)= - \left( \frac{3v^2 {\cal{H}}^2}{ V^2 \kappa^2}+\frac{R}{2 \kappa^2} \right)  \frac{\phi_0^2}{2}+ \frac{v^2J_0 }{V^2\kappa^2}\phi_0,
\end{equation}
where $R$ is the Ricci scalar. The factor $\kappa$ is introduced because Eq.\eqref{effective_eom_scalarfield_background} is written in terms of $x^0$, which has mass dimension $1$, whereas the action above is written in terms of $X^0$, which has mass dimension $-1$. In this way, the resulting modified scalar-field dynamics remain compatible both with the classical late-time metric and with the quantum-gravity-reconstructed geometric sector.

\

\label{Sec:Perturbations}

\textbf{\textit{Effective field theory of cosmological perturbations}} --- 
Following the same procedure used for the homogeneous sector, we now derive the effective dynamics of the scalar field perturbations. As in standard cosmological perturbation theory, perturbations of the emergent geometry induce corresponding modifications in the matter dynamics \cite{MUKHANOV1_action_perturbation,Branden_perturb, Tapia2020}.
The equations for the condensate density and phase perturbations can be written as
\begin{align}
     \left(\tilde\Box-\eta\right)\delta\rho=\mathcal{D}\left[\delta\theta\right]\, ,\\
\label{EQ_EOMPerturb_theta}
\tilde\Box\delta\theta+2\frac{\dot\rho_0}{\rho_0}\delta\dot\theta+\mathcal{L}\Big[\left(\tilde\Box-\eta\right)^{-1}\mathcal{D}[\delta\theta]\Big]&=0\,,
\end{align}
where we have defined the following differential operators
\begin{align}
    \label{Modified_box}
    \tilde\Box &=\partial_0^2+\alpha_r\nabla^2\,,\\
    \label{EQ_densityPT_eq}
     \mathcal{D}&=\rho_0\left[ \left(2\dot\theta_0-\gamma\right)\partial_0-\alpha_i\nabla^2 \right] \,,\\
     \label{operator_diff_L}
    \mathcal{L}&=\rho_0^{-1}\left[ \left(2 \dot\theta_0-\gamma\right)\partial_0+\alpha_i\nabla^2+\ddot\theta_0-\beta\right]\, ,
\end{align}
and the explicit equation of motion for the phase perturbation is given by
  \begin{widetext}
\begin{equation} \label{eq:full_eom_delta_theta}
\begin{split}
&\underbrace{\Bigl[ 1
+
\mathcal Q(2\dot\theta_0-\gamma)^2\Bigr] }_A
\delta\ddot\theta+ \underbrace{\Bigl[
2 \frac{\dot\rho_0}{\rho_0}
+\frac{2\dot\theta_0-\gamma}{\rho_0} 
\mathcal Q\!\bigl[\partial_0\!(\rho_0 (2\dot\theta_0-\gamma))\bigr]
+\mathcal Q\left(\ddot\theta_0-\beta \right)
\left(2\dot\theta_0-\gamma\right)
\Bigr] }_B
\delta\dot\theta+\\
&\underbrace{+\Bigl[
\alpha_r
-\alpha_i\mathcal Q\left(\ddot\theta_0-\beta\right) 
+\frac{2\dot\theta_0-\gamma}{\rho_0} 
\mathcal Q\!\bigl[-\alpha_i \dot \rho_0 \bigr]
\Bigr] }_C
\nabla^2\delta\theta
\underbrace{-\alpha_i\mathcal Q\left(2\dot\theta_0-\gamma\right)}_D
\partial_0\nabla^2\delta\theta
 \underbrace{-\alpha_i^2 \mathcal Q}_E \nabla^4\delta\theta
 = 0\,.
\end{split}
\end{equation}
\end{widetext}
where, for notational convenience, we introduced the inverse operator  $\mathcal{Q}=\left(\tilde\Box-\eta\right)^{-1}$. 
When $\alpha_i=0$, which also implies $\beta=0$, the equation of motion for the phase perturbation decouples from $\delta \rho$ and reduces to 
\begin{equation}
    \label{eq_ag_phase_decoupled}
\lambda_1(x^0,\pi_\phi)\delta \ddot{\theta}+\lambda_2(x^0,\pi_\phi)\delta \dot{\theta}+\alpha_r \nabla^2 \delta \theta=0\,,
\end{equation}
where we have defined the following time-dependent coefficients 
\begin{equation}
\begin{split}
\lambda_1&=\mathcal{Q}\left[1+\left(2\dot \theta_0-\gamma\right)^2\right]\,,\\
\lambda_2=& \mathcal{Q}\left[2+\left(2 \dot{\theta}_0-\gamma\right)^2\right] \frac{\dot{\rho}_0}{\rho_0}+3\mathcal{Q}\left(2 \dot{\theta}_0-\gamma\right) \ddot{\theta}_0\,.
\end{split}
\end{equation}


Perturbing the expectation value of the scalar-field operator, $\delta\phi=\delta\langle\hat{\Phi}\rangle_\sigma$, we have
\begin{equation}
\label{eq:delta_phi_of_d_pi_delta_theta}
\delta\phi= 2 \rho_0\,\delta \rho\, \partial_{\pi_\phi} \theta_0+ \rho_0^2 \, \partial_ {\pi_\phi}\delta \theta\,.
\end{equation}
\paragraph{Scalar field dynamics in the coupled regime.}
To extract an effective equation of motion for the perturbations of the scalar field, we follow the same line of reasoning used for the homogeneous sector. Differentiating the full equation of motion for the perturbed phase given in Eq.\eqref{eq:full_eom_delta_theta} with respect to $\pi_\phi$ leads to an equation of motion for $\partial_{\pi_\phi} \delta \theta$  containing source-like terms generated by derivatives acting on the background-dependent coefficients.
In the early-time regime, the operator ${\cal{Q}}$ can be approximated as ${\cal{Q}} \simeq q \mathbb{I}$, allowing the coefficient operators in the above equation to be treated as standard functions. This replacement mirrors the standard step in BEC analogue gravity, where the inverse kinetic operator generated by the density-phase inversion is rendered a multiplicative coefficient by neglecting its differential part against the dominant algebraic one, as appropriate for long-wavelength perturbations \cite{Torrome:2015cga_Liberati, Analogue_Gravity, Barcelo:2001tb}.  Now, substituting the inverted relation for $\partial_{\pi_\phi} \delta \theta$ as a function of $\delta \phi$ (obtained from Eq.\eqref{eq:delta_phi_of_d_pi_delta_theta}) 
 into the perturbation equation Eq.\eqref{eq:full_eom_delta_theta} yields the effective equation of motion for the scalar-field perturbation
\begin{equation}
    \label{eq:eom_perturbations_delta_phi}
    \delta \ddot \phi+ f_1\, \delta \dot \phi+ f_2\, \nabla^2 \delta \dot \phi+f_3\, \nabla^2 \delta \phi+ f_4\,\nabla^4 \delta\phi+ f_5\, \delta \phi+ \tilde J=0\,,
\end{equation}
where the coefficients $f_i$ are (complicated) functions of the background dynamical variables (volume and scalar field), determined by the background condensate dynamics, depending on $x^0$ and $\pi_\phi$
\begin{equation}
\begin{split}
\label{eq:modfied_klein_gordon_coefficients}
    f_1&= \frac{B}{A}-6 {\cal{H}}\,,\quad f_2=\frac{D}{A}\,,\quad 
    f_3 =\frac{C}{A}- 3 {\cal{H}}\, \frac{D}{A}, \\
    f_4 &= \frac{E}{A}\,,\quad  f_5 = 3 \Big(  3 {\cal{H}}^2- \frac{B}{A} \,{\cal{H}}-\dot{{\cal{H}}}\Big)\,.
    \end{split}
\end{equation}
Here we have isolated the dependence on the Hubble rate $\mathcal{H}$, and $\tilde J(x^\mu,\pi_\phi)$ denotes a nontrivial source term depending both explicitly on the spacetime coordinates and implicitly through the perturbations of both the volume  $\delta V$ and $\delta\theta.$ 

The resulting equation therefore departs significantly from the standard Klein–Gordon form, entailing higher-derivative contributions such as $\nabla^2\delta\dot\phi$ and $\nabla^4\delta\phi$, which encode genuine quantum-gravity corrections arising from the underlying quantum dynamics of the GFT condensate.
Nevertheless, these two terms are proportional to $\alpha_i$ and $\beta$ and are therefore suppressed.

As in the homogeneous sector, one may reinterpret this effective dynamics in terms of an emergent classical field theory, either by modifying the effective geometry experienced by the scalar field, the kinetic structure of the field itself, or both in parallel. 



\paragraph{Scalar field dynamics in the decoupled regime.}
If we now assume the decoupling regime ($\alpha_i=\beta=0$), the coefficients in Eq.\eqref{eq:modfied_klein_gordon_coefficients}  can be evaluated explicitly in both the early and late-time limits. One obtains, respectively,
\begin{equation}
\begin{aligned}
    \delta \ddot \phi  -& 3 {\cal{H}}\,\frac{1+ 12\,c^2\, a^{-6} }{1+ 4\,c^2\, a^{-6}} \, \delta \dot \phi+ 3 \left(24 \,{\cal{H}}^2\, \frac{c^2\,a^{-6}}{1+4\,c^2\, a^{-6}}- \dot {\cal{H}}  \right) \delta \phi\\
    &+\frac{\alpha_r}{1+4\,c^2\, a^{-6}} \nabla^2 \delta \phi + \tilde J =0\,,
    \end{aligned}
\end{equation}
and 
\begin{equation}
    \delta \ddot \phi  - 3 {\cal{H}}\,\, \delta \dot \phi - 3  \dot {\cal{H}}\, \delta \phi+ \alpha  \nabla^2 \delta \phi + \tilde J =0\,,
\end{equation}
with $a:= (V/v)^{1/3}$ being the scale factor.
As expected, the late-time equation is recovered directly from the early-time dynamics in the limit $a\gg 1$, where the quantum-gravity corrections become negligible. We see that the genuinely nonstandard contributions arise only in the fully coupled regime of density and phase perturbations.

\paragraph{Scalar field dispersion relation.}
To derive the explicit form of the dispersion relation, we consider Eq.\eqref{eq:eom_perturbations_delta_phi} neglecting $\tilde J$, and we perform a spatial Fourier transform $\delta \phi(x^0,x^i)= \int {\rm{d}}^3 k(2 \pi)^{-3}\,\delta \phi_k(x^0)\, {\rm{e}}^{i \,k_i\,x^i}$.
Assuming an adiabatic regime, we further adopt a WKB ansatz for the time-dependent modes 
$\delta\phi_k(x^0) \simeq (2 \Omega_k(x^0))^{-1/2}\, {\rm{e}}^{-i \int^{x^0} {\rm{d}} s\, \Omega_k(s)}$, under  the adiabatic condition $\left|\dot{\Omega}_k(x^0)/\Omega_k(x^0)^2 \right| \ll 1$.
Substituting the above expression into the homogenous version of Eq.\eqref{eq:eom_perturbations_delta_phi} yields the effective dispersion relation
\begin{equation}
\begin{split}
    \Omega_k(x^0) &= \frac{i}{2} \Gamma_k(x^0) \,\\
    &\pm \,\sqrt{\Gamma_k(x^0)^2+ 4 \left[  k^4\, f_4(x^0)-k^2\,f_3(x^0)+f_5(x^0) \right]},
    \end{split}
\end{equation}

where we have defined $\Gamma_k(x^0):= k^2 \,f_2(x^0)-f_1(x^0)$.
The resulting dispersion relation exhibits both dissipative and dispersive features.  In the standard GR case, for certain gauge choices (for instance, cosmic time), imaginary terms also appear in the dispersion relation. These terms are associated with the so-called Hubble friction, and they do not yield a genuinely dissipative behavior, as they are $k$-independent. In our case, we also find a term resembling Hubble friction, but in addition, we have a $k$-dependent term that signals a genuinely (albeit emergent, non microscopic) dissipative behavior \cite{Clifton2012,Liberati2014}. 

Furthermore, we observe dispersive effects arising from the real part of $\Omega_k$. In fact, the presence of these terms in the dispersion relation of the scalar field is simply an indicator of the effective approach we are adopting for the dynamics. However, analogously to the background case, the $-3 {\cal{H}}  \delta \dot \phi$ term also produces modes growing in time proportionally to the volume.  As these modes are unphysical, they must be removed by restricting the solution space through the additional requirement that the scalar field be an intensive quantity.
\textbf{\textit{Conclusion}} --- 
We have derived an effective scalar-field theory dynamics from (relational) GFT condensate cosmology, using the collective hydrodynamics of the condensate as the starting point rather than assuming a background spacetime from the outset. Therefore, the derivation is fully embedded within the fundamental quantum gravity formalism, and allows to study its physical implications. At the homogeneous level, the scalar field reconstructed from the condensate phase obeys a modified equation of motion on the emergent cosmological background. In the late-time regime, where the condensate geometry approaches a spatially flat FLRW spacetime in harmonic time, this dynamics reduces to the expected massless-scalar behavior, while deviations persist in the high-curvature regime near the bounce.

We then extended the construction to scalar perturbations, i.e. inhomogeneities, also reconstructed fully from within the fundamental quantum gravity formalism (in its mean field approximation). The resulting effective wave equation is controlled by background-dependent coefficients fixed by the underlying GFT condensate. In the decoupled regime, the perturbations reproduce the expected relativistic form in the late-time limit. In the coupled density-phase regime, however, the effective dynamics contains mixed time-space derivatives and higher spatial derivatives. These terms lead to a modified dispersion relation exhibiting both dispersive and effective dissipative contributions. They are therefore direct signatures, at the level of emergent matter propagation, of the microscopic quantum-geometric structure encoded in the condensate.

Our results show that GFT condensate cosmology does not only provide an emergent background geometry, but also determines the effective dynamics of matter fields propagating on it. This opens a concrete route toward extracting phenomenological consequences of quantum geometry from GFT quantum gravity, especially through departures from standard relativistic propagation in the early universe. Future work should clarify the role of the source terms, the precise status of the intensive scalar perturbation, and the connection between the modified dispersion relation derived here and observable cosmological perturbation spectra.

\textbf{\textit{Acknowledgments}} --- 
DO acknowledges support from Grant PR28/23 ATR2023-145735 funded by MCIN/AEI/10.13039/501100011033. RD acknowledges support from the COST action 23130 (\lq BridgeQG\rq). FG acknowledges support from the Istituto Nazionale di Fisica Nucleare (INFN) through
the Theoretical Astroparticle Physics (TAsP) project, from the Dipartimento di Fisica e Astronomia (DFA) of the University of Padua and from WOST, WithOut SpaceTime project (https://withoutspacetime.org), supported by Grant ID 63683 from the John Templeton Foundation (JTF).




\bibliography{biblio}

\end{document}